\documentclass[11pt,tightenlines,twoside,preprintnumbers,showpacs,showkeys]{revtex4}
\usepackage[english,USenglish]{babel}
\usepackage{amsmath}
\usepackage{amssymb}
\usepackage{psfrag}
\usepackage{graphicx}
\usepackage{dsfont}
\usepackage{bibgerm}
\usepackage{url}
\usepackage[dvips,letterpaper,nofoot]{geometry}
\usepackage[T1]{fontenc}

\usepackage{mathrsfs}
\DeclareSymbolFontAlphabet{\mathrsfs}{rsfs}

\DeclareMathAlphabet{\mathcal}{OMS}{cmsy}{m}{n}

\newcommand{\scri}{\mathrsfs{I}}
\newcommand{\Scri}{\scri}

\newcommand{\izero}{\imath^0}
\newcommand{\const}{\text{const}}

\newcommand{\labelsize}{\tiny\fontsize{1.2em}{0pt}\selectfont\strut}
\newcommand{\llabelsize}{\tiny\fontsize{1.8em}{0pt}\selectfont\strut}

\newcommand{\PSFig}[4]{\resizebox*{#1}{#2}{\rotatebox{#3}{\includegraphics{#4}}}}

\newcommand{\A}{\alpha}
\renewcommand{\S}{\beta}
\renewcommand{\a}{\chi}
\newcommand{\K}{\chi}
\newcommand{\g}{\Gamma}

\newcommand{\eps}{\epsilon}
\newcommand{\del}{\partial}
\newcommand{\E}{\mathcal{E}}%
\newcommand{\B}{\mathcal{B}}%

\newcommand{\mf}{\mathcal{M}}
\newcommand{\mft}{\tilde{\mf}}

\begin{document}

\title{Hyperboloidal data and evolution}

\author{S. Husa}
\email{sascha.husa@uni-jena.de}
\affiliation{Theoretisch-Physikalisches Institut, University of Jena,
             Max-Wien-Platz 1, D-07743 Jena, Germany}

\author{C. Schneemann}
\email{carsten.schneemann@aei.mpg.de}
\author{T. Vogel}
\email{tilman.vogel@aei.mpg.de}
\author{A. Zengino\u{g}lu}
\email{anil.zenginoglu@aei.mpg.de}
\affiliation{Max-Planck-Institut f\"ur Gravitationsphysik,
             Albert-Einstein-Institut, 
             Am M\"uhlenberg 1, D-14476 Potsdam, Germany}

\begin{abstract}
  We discuss the hyperboloidal evolution problem in general relativity
  from a numerical perspective, and present some new results.
  Families of initial data which are the hyperboloidal analogue of
  Brill waves are constructed numerically, and a systematic search for
  apparent horizons is performed.  Schwarzschild-Kruskal spacetime is
  discussed as a first application of Friedrich's general conformal
  field equations in spherical symmetry, and the Maxwell equations are
  discussed on a nontrivial background as a toy model for continuum
  instabilities.
\end{abstract}



\preprint{AEI-2005-183}

\maketitle

%%%%%%%%%%%%%%%%%%%%%%%%%%%%%%%%%%%%%%%%%%%%
%% MAINMATTER
%%%%%%%%%%%%%%%%%%%%%%%%%%%%%%%%%%%%%%%%%%%%
\section{Introduction}

In this paper we consider algorithms for numerical relativity (NR) based on 
hyperboloidal slices -- spacelike hypersurfaces characterized
by a mean extrinsic curvature $\chi$ that approaches a finite value in
the limit $r \rightarrow \infty$. Correspondingly such slices are not 
asymptotically euclidean, but rather reach out to null infinity, and thus
provide an alternative to null surfaces for 
tracking radiation signals to large distances from their source, e.g.
in order to predict signals in a gravitational wave detector
(see e.g. \cite{Joerg:LivRev, Husa:2002zc}).
Being spacelike, hyperboloidal slices are in some sense more flexible than 
null surfaces, and thus  interesting for constructing numerical relativity codes aimed at 
gravitational wave physics.
In this paper we briefly discuss our general ideas about the design of numerical
codes for hyperboloidal evolution and some preliminary results from
two perspectives we believe to be of key importance: the need to
control continuum instabilities and fitness to accurately resolve
gravitational wave signals. We also list a few new results.

As is the case with other disciplines of computational physics, an essential part of the art of NR is to make the physical
continuum features manifest in the discrete system that is then solved by a computer.
In general relativity (GR) diffeomorphism invariance gives rise to a variety of problems
not familiar from other theories and which are not yet understood in sufficient depth 
to provide a fully satisfactory basis for numerical simulations.
A typical evolution scheme has many more computational than physical degrees of freedom, the extra degrees of freedom correspond to gauge choice and the presence of constraints -- one should therefore not be surprised to find a generic tendency for instabilities in the excess degrees of freedom, and indeed the multitude of formulations of the Einstein equations are typically plagued by
instabilities whose precise causes have often remained elusive, and we expect
much further work to be necessary in order to understand the dos and don'ts
of NR. As a simple (linear) example for the type of problems that have
to be expected, we will consider hyperboloidal evolution of an
electromagnetic field on Minkowski background.

At least from an observational point of view it is clearly desirable to design
numerical codes with accurate GW signal prediction in mind.
This is difficult for various reasons. First, in all physically relevant scenarios gravitational radiation is only a relatively small effect in the energy balance of the system. Second, in GR such fundamental quantities
as energy, momentum, or emitted gravitational radiation energy can only
be defined unambiguously in terms of asymptotic limits. Consequently, it also becomes
particularly difficult to formulate physically motivated boundary conditions
along the lines of ``outgoing radiation boundary conditions'' at finite 
distance from the sources. 

Conformal compactification, originally suggested by Penrose \cite{Penrose:1963}
allows to discuss asymptotics in terms of local differential geometry and 
has provided a very fruitful framework to approach many problems in mathematical
relativity. Naturally, it also raises hopes for a consistent notion and quantitative
treatment of GW signals. However, since the conformal framework is extremely flexible,
it does not by itself determine a strategy for NR, and
additional physical intuition and practical insights are necessary to bring this
technique to fruition in numerical simulations. In the following, we will briefly review
the connection between asymptotics, conformal 
compactification and gravitational wave (GW) signals 
before sketching our strategy to develop codes for the hyperboloidal evolution problem.
We then present some new numerical results concerning 
Friedrich's general conformal field equations in spherical symmetry as a simple 
window into the interplay of spatial and null infinity, 
the Maxwell equations on a nontrivial background as a toy model for continuum instabilities,
and initial data that generalize Brill waves to the hyperboloidal context.

\section{Conformal compactification and rescaling}
A key idea behind conformal rescaling is to compute ``order unity'' quantities, e.g. for a massless scalar field $\Phi$ a rescaling of the type $\Psi:= r\, \Phi$, which asymptotically just cancels the known fall-off of the radiation from an isolated source. This allows one to work with quantities that are finite even asymptotically.
Such a procedure can furthermore improve the numerical conditioning of radiation problems.  Generally, it is useful in computational work to factor out what is already known. The idea of conformal compactification is to perform a conformal transformation on the metric
$g_{ab}= \Omega^2 \tilde g_{ab}$ and view 
the physical space-time $\mft$ as a submanifold $\mft = \{ p \in \mf \, \vert \, \Omega(p) > 0  \}$ 
of some manifold $\mf$ completed by boundary points 
$\partial \mft = \{ p \in \mf \, \vert \, \Omega(p)= 0  \}$ lying ``at infinity'' with 
respect to $\tilde g_{ab}$.
The definition of a certain type of asymptotics, like asymptotic flatness, then
proceeds in terms of asymptotic properties of the conformal factor, which define
a desired physical fall-off behavior (see e.g. \cite{Joerg:LivRev}).
Note that in a relativistic theory, we need to deal with three types of directions toward infinity: timelike ($\imath^\pm$), spacelike ($\izero$) and null ($\scri^\pm$), and these limits have very different physical significance. In particular, observers situated at ``astronomical'' distances (e.g. GW detectors) can be modeled through geometric objects at future null infinity \cite{Joerg:98rad}.
Clearly, a thorough physical understanding of the problem of consistently modeling GW sources and detectors in a single picture is very desirable.

However, writing the Einstein tensor in terms of the rescaled metric makes 
it immediately clear that taking this concept to the level of the field
equations can not be straightforward:
$$
\tilde G_{ab}[\Omega^{-2} g] =  G_{ab}[g] - \frac{2}{\Omega}
\left(\nabla_a \nabla_b\Omega - g_{ab} \nabla_c \nabla^c \Omega \right)
 -\frac{3}{\Omega^2} g_{ab}\left(\nabla_c \Omega \right)\nabla^c \Omega.
$$
In the new variables the equations are formally singular at $\partial\mft$ 
whereas multiplication by $\Omega^2$ leads to a degenerate principal part for $\Omega=0$. 
A very general prescription for regularizing the rescaled Einstein equations
has been obtained by Friedrich through the formulation
of the regular conformal field equations \cite{Friedrich:CFEs}.
The fact that this is actually possible for the Einstein equations,
is a nontrivial result and may certainly seem surprising.
Unfortunately, it is achieved at a high price of introducing a large number of new evolution 
variables, which complicates the numerical implementation and increases the risk of triggering 
continuum instabilities (for numerical results see \cite{Husa:2002zc}). 

Compactification techniques have been used in NR for quite some time, but have
often been based on less general regularization techniques, e.g. through restriction to a special class of gauges. 
Compactification in null directions has been very successful in the characteristic approach (see e.g. \cite{Bishop:1997ik} and \cite{Husa:2003mc} for recent results) and is well understood. 
Compactification of spacelike infinity has not only been used to construct initial data 
(see e.g. \cite{Husa:1996xz, Brandt:1997tf, HusaDiss}), but encouraging results have also been obtained in the time evolution problem
\cite{Bernd:1997}, where black holes are modeled as ``internal asymptotic ends'', often referred to as punctures, and recently also to get rid of the boundary problem in NR \cite{Pretorius:2005prl}. In the evolution context, however, some open questions remain, e.g. because compactification at $\izero$ leads to a ``piling up'' of waves. At $\scri^+$ this effect does not appear -- waves leave the physical spacetime through the boundary $\scri^+$. Also, regularity issues of the equations at spatial infinity are not yet fully understood, although much progress has been made with Friedrich's general conformal field equations \cite{Friedrich:Cargese}, for which we discuss a simple application below.

\begin{figure}[htbp]
  \centering
  \psfrag{EH}{\llabelsize EH}
  \psfrag{Scri}{\llabelsize $\Scri$}
  \psfrag{null}{\labelsize null}
  \psfrag{geodesics}{\labelsize geodesics}
  \psfrag{excised}{\labelsize excised}
  \includegraphics[width=0.49\textwidth,height=5cm]{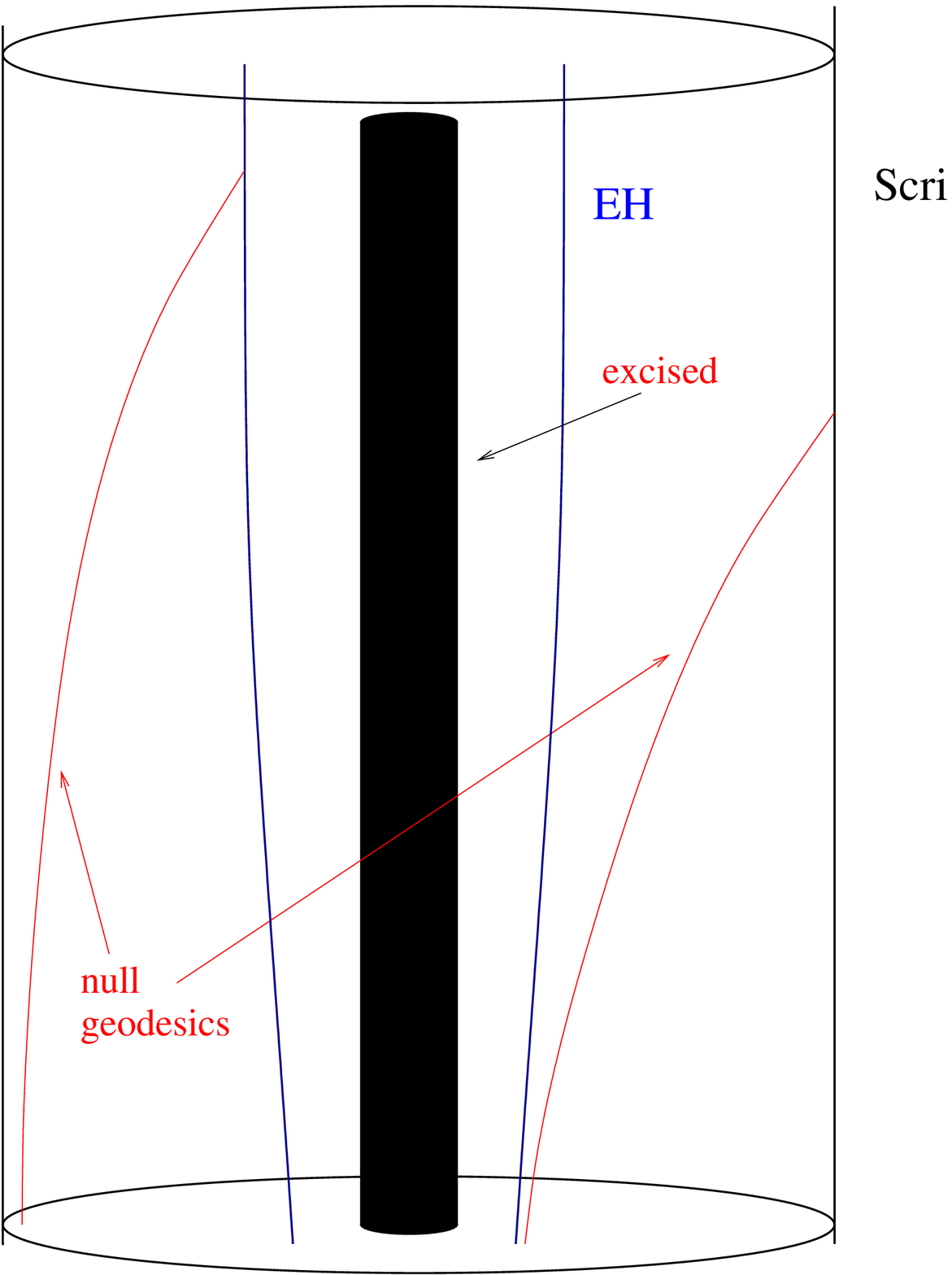}
  \includegraphics[width=0.49\textwidth]{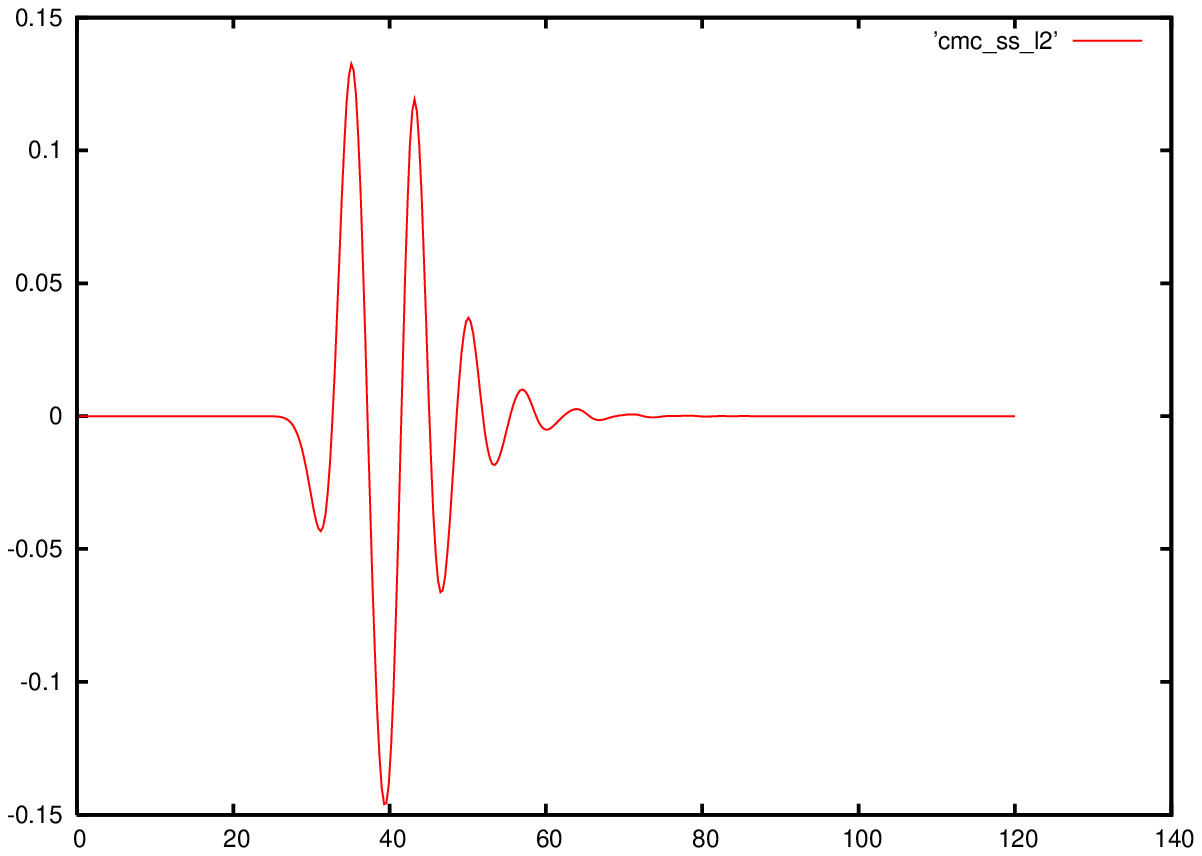}
  \caption{Left: A sketch depicting a situation with $\scri^+$ frozen to a coordinate sphere. Right: The ringdown of a scalar field with angular momentum number $l=2$ on a compactified Schwarzschild background with CMC-slices and $\scri^+$ frozen to a coordinate sphere. \label{fig:scrifix}}
\end{figure}

A code that utilizes hyperboloidal slices to compactify null 
infinity can profit from all the flexibility in gauge that a Cauchy approach offers. However, following the idea to factor out what is already known and making the physical continuum features also manifest in the
numerical code leads to the problem of making manifest the rigid structure of
null infinity in addition to the fall-off of the ``gravitational field''.
Particularly important seem the shear-free property of null infinity and
the existence of a natural class of time coordinates associated with
affine parameters of the null geodesic generators of $\scri$, known as 
Bondi time. It is this time coordinate which corresponds to the proper time
of distant observers \cite{Joerg:LivRev}, and which thus corresponds to an 
``undistorted signal'', as in Fig. \ref{fig:scrifix}.
We suggest to use the gauge freedom to make the rigid structure of $\scri$
manifest and freeze it to a fixed coordinate sphere as discussed in detail by
Andersson \cite{Andersson:2002tue} (in particular here the connection between 
the 3+1 split and the Bondi gauge is discussed, and essentially the same
recommendation to use such a gauge as starting point for regularization
is given).
Fixing $\scri^+$ to a coordinate sphere, it is natural to identify it also
with the boundary of the computational domain, and thus
to restrict oneself to the physical part of the spacetime. 
First experiments along these lines with scalar fields on a Schwarzschild background
have yielded the ringdown results in Fig. \ref{fig:scrifix}.

As an example for the type of coordinates we have in mind,
consider computing just the domain of dependence of a piece of Minkowski space
with initial data given on a ball. Appropriate coordinates are those
which are also adapted to self-similarity:
$$
ds^2 = \frac{e^{-2 \tau}}{R^2}\left[ -(R^2 - r^2) d\tau^2 - 2 r dr d\tau
  + dr^2 + r^2 \left(d\theta^2 + \sin{\theta}^2 d\varphi^2 \right) \right].
$$
Using the same type of coordinates in the compactified spacetime with $R$ identified
with the initial location of $\scri$ yields the picture in fig. \ref{fig:scrifix}.
Freezing $\scri$ to a coordinate sphere essentially corresponds to a choice 
of the shift vector on  $\scri$, which leads to two problems:
First, the prescription of shift  needs to be
compatible with a well--posed evolution system, and second, 
one also needs to choose well for the shift vector away from $\scri$,
in order not to distort the geometry in the interior of the spacetime.

Our strategy to develop codes for the hyperboloidal initial value problem has thus been 
threefold: First, we have developed a computational infrastructure that allows us to confront
equations as complex as the conformal field equations without tying us down to a particular
form of the equations.
To this end, the {\tt Kranc} code generation and tensor manipulation package \cite{Husa:2004ip} has been developed.
Second, it has proven very fruitful to learn as much as possible from the characteristic approach,
which is less general, but works well.
Third, we have started a number of smaller projects that allow us test what works and what does
not in simplified situations, and actually start a mathematical analysis of the properties of our algorithms.
We present some preliminary results below.

\section{General conformal field equations}

The approach suggested above and sketched in Fig. (\ref{fig:scrifix}) is well adapted
to computing gravitational wave signals, but can not reproduce a global representation
of the spacetime, which includes spacelike infinity $\izero$. From the point of view
of an observer at $\scri^+$, $\izero$ represents the infinite past, which is clearly 
relevant for certain questions, e.g. a quasi-stationary solution may have persisted for
a very long time, before violent dynamics sets in.
\begin{figure}[btp]
\centering
\psfrag{T}{\llabelsize$\tau$}
\psfrag{i+}{\labelsize$\imath^+$}
\psfrag{J+}{\labelsize$\scri^+$}
\psfrag{I}{\labelsize$\mathcal{I}$}
\psfrag{d/M}{\labelsize d/M}
\psfrag{2}{\labelsize 2}
\psfrag{10}{\labelsize 10}
\psfrag{300}{\labelsize 300}
\psfrag{ingoing}{\labelsize ingoing}
\psfrag{outgoing}{\labelsize outgoing}
\psfrag{r=const}{\labelsize r = const}
\psfrag{singularity}{\labelsize singularity}
\psfrag{horizon}{\labelsize horizon}
\includegraphics[width=0.6\textwidth]{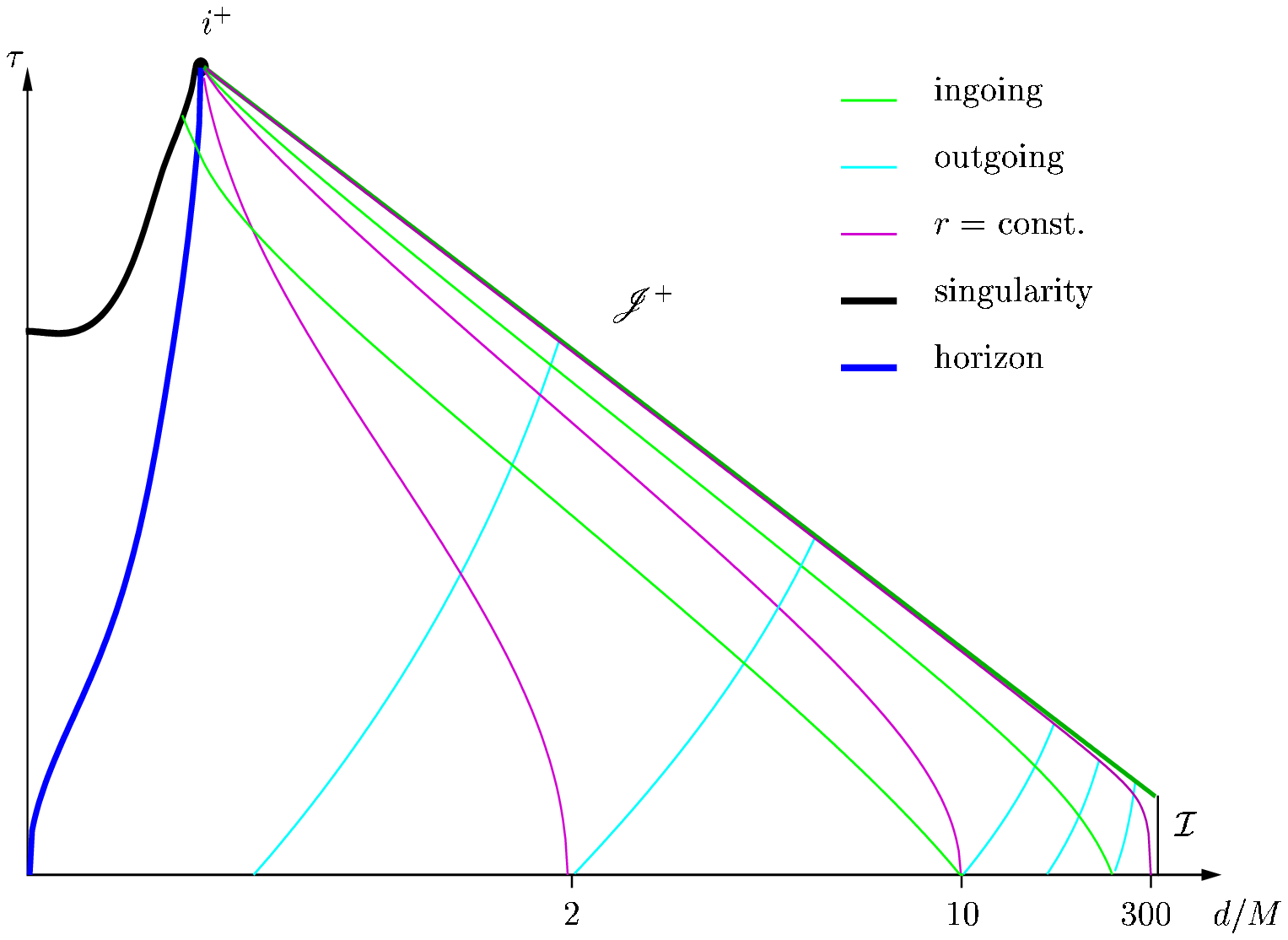}
\caption{Schwarzschild-Kruskal spacetime in a conformal Gauss gauge.\label{fig:conf_geodesics}}
\end{figure}
Friedrich's general conformal field equations \cite{Friedrich:Cargese}, which rely on 
the conformal Gauss gauge, allow for a global treatment, in which different asymptotic regions can be handled with one system of equations. 
Using this method can provide initial data for a hyperboloidal code that is actually determined
from a Cauchy surface. Also, numerical experiments with this system might give rise to a better understanding of the regularity issues around spatial infinity.

As a first step we have used the general conformal field equations to construct the 
Schwarzschild-Kruskal solution with initial data specified on a Cauchy surface. Using 
the conformal Gauss gauge, in which by spherical symmetry all equations become 
ordinary differential equations, it was possible for the first time to cover the 
entire Schwarzschild-Kruskal spacetime including spacelike, null and timelike 
infinity and the domain close to the singularity (Fig. \ref{fig:conf_geodesics}). 
These results can also be seen as a feasibility study of the conformal Gauss gauge. 
Current work is directed to the numerical solution of the general conformal field 
equations for non-spherically symmetric initial data.

\section{Minkowski space}

A natural first exercise when entering uncharted territory in NR is to
consider Minkowski space. 
A comparison of evolutions of Minkowski space in various gauges with
the full conformal field equations has been reported in \cite{Husa:2002zc}, a
particularly interesting case is, when $\scri^+$ is identified with a
fixed coordinate sphere and the conformal geometry is chosen
stationary. In this case a constraint violating continuum instability
is found. Inspection of the equations suggests the instability to be
due to the type of effect described in this section.  Recently, the
mathematical tools to clarify this have been discussed by Frauendiener
and Vogel \cite{Tilman:2004}.

As a simple exercise for this type of problem, we have analyzed the
case of a Maxwell field ($E^a$, $B^a$) on Minkowski space sliced by
non-trivial hypersurfaces:
\begin{align*}
  \del_t E^a
  - \S^b \del_b E^a
  - \A\, \eps^{abc} h_{cd} \del_b B^d &
  =
    \alpha (
      \eps^{abc} h_{cd} \a_b B^d
    - \K E^a
    + \eps^{abc} h_{cd} \g^d_{eb} B^e
  ) 
  - E^b \del_b \S^a\,,\\
%%%%%%%%%%%%%%%%%%%%%%%%%%%%%%%%%%%%%%%%%%%%%%%%%%%%%%%%%%%%%%%%%%%%%%
  \del_t B^a
  - \S^b \del_b B^a
  + \A\, \eps^{abc} h_{cd} \del_b E^d &
  = -\alpha (
     \eps^{abc} h_{cd} \a_b E^d
    + \K B^a
    + \eps^{abc} h_{cd} \g^d_{eb} E^e
 ) 
  - B^b \del_b \S^a
\end{align*}
The constraints and constraint propagation equations then are
\begin{align*}
  0 = \E &:=
    \del_b E^b
  + \g^b_{cb} E^c, &   \del_t \E - \S^a\del_a\E &= - \A \, \K \E
  \,,\\
%%%%%%%%%%%%%%%%%%%%%%%%%%%%%%%%%%%%%%%%%%%%%%%%%%%%%%%%%%%%%%%%%%%%%%
  0 = \B &:=
  \del_b B^b + \g^b_{cb} B^c, & \del_t \B - \S^a\del_a\B &= - \A \, \K \B\,.
\end{align*}
Here, $h_{ab}$, $\chi$, $\chi_a$, $\g^a_{bc}$, $\alpha$ and $\beta^a$
are 3-metric, mean extrinsic curvature, acceleration, Christoffel
symbol of $h_{ab}$, lapse and shift respectively. From the constraint
propagation equations one directly reads off a stability prognosis in
the spirit of Frauendiener and Vogel \cite{Tilman:2004}: if $\chi <
0$, a constraint violating continuum instability has to be expected,
whereas $\chi > 0$ should result in constraint damping.  Both effects
will be demonstrated below for a simple class of slices in Minkowski
space with a fixed sign of $\chi$.
Note that densitizing the evolved fields can change the sign of the
$\chi$ factors. This example thus demonstrates that one needs to be
aware of a subtle interplay between the evolution system, choice of
variables and gauge.
Continuum instabilities of this type are essentially an ODE effect
in the sense that they are determined by lower order source terms rather than
spatial derivatives.
Consequently, it is important to realize that for numerical purposes, analyzing 
the principal part is only a starting point. In general, lower order terms have 
to be carefully analyzed and a formulation of the theory has to be chosen that avoids instabilities.
Clearly, this process benefits from avoiding excess baggage when formulating the equations one starts with. 
\begin{figure}[t]
  \centering
   \includegraphics[width=0.49\textwidth,height=4.9cm]{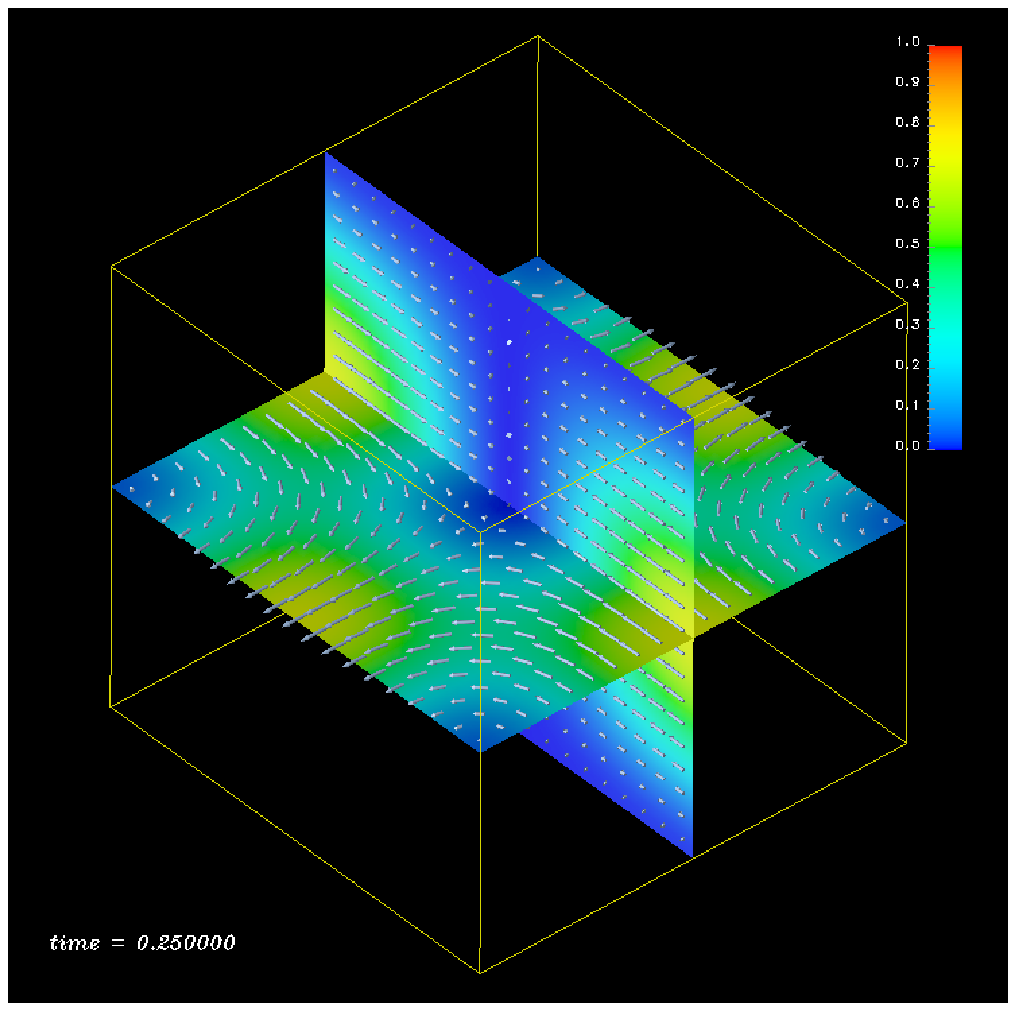}
   \hfill
   \includegraphics[width=0.49\textwidth]{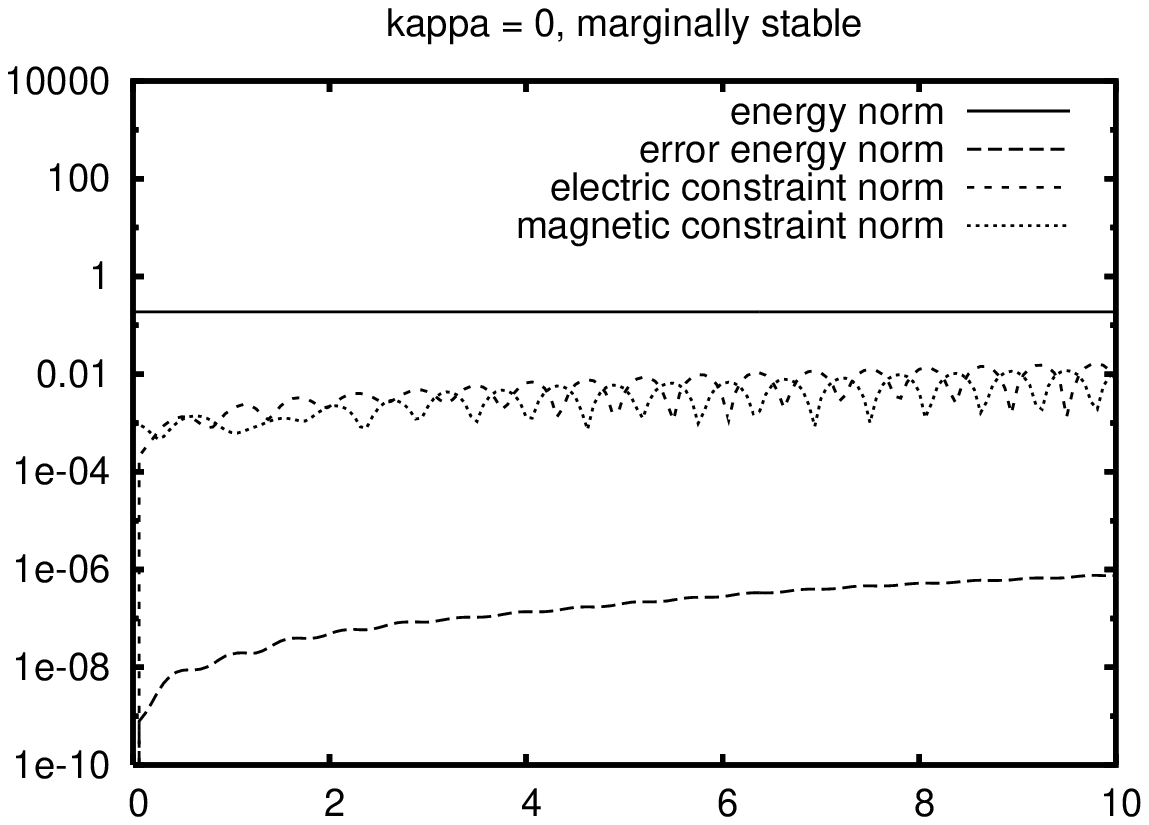}
  \caption{%
    (a) The electric field configuration and energy density.
    (b) Evolution in standard Minkowski coordinates;
    behaves nicely, energy conserved, linear drift away from exact
    solution (error energy norm depends linearly on time). Oscillations
    in the constraints due to lowered accuracy of constraint calculation
    at the boundary (stencil limitation).%
    \label{fig:minkcomp1}
  }
\end{figure}
\begin{figure}[b]
  \centering
   \includegraphics[width=0.49\textwidth]{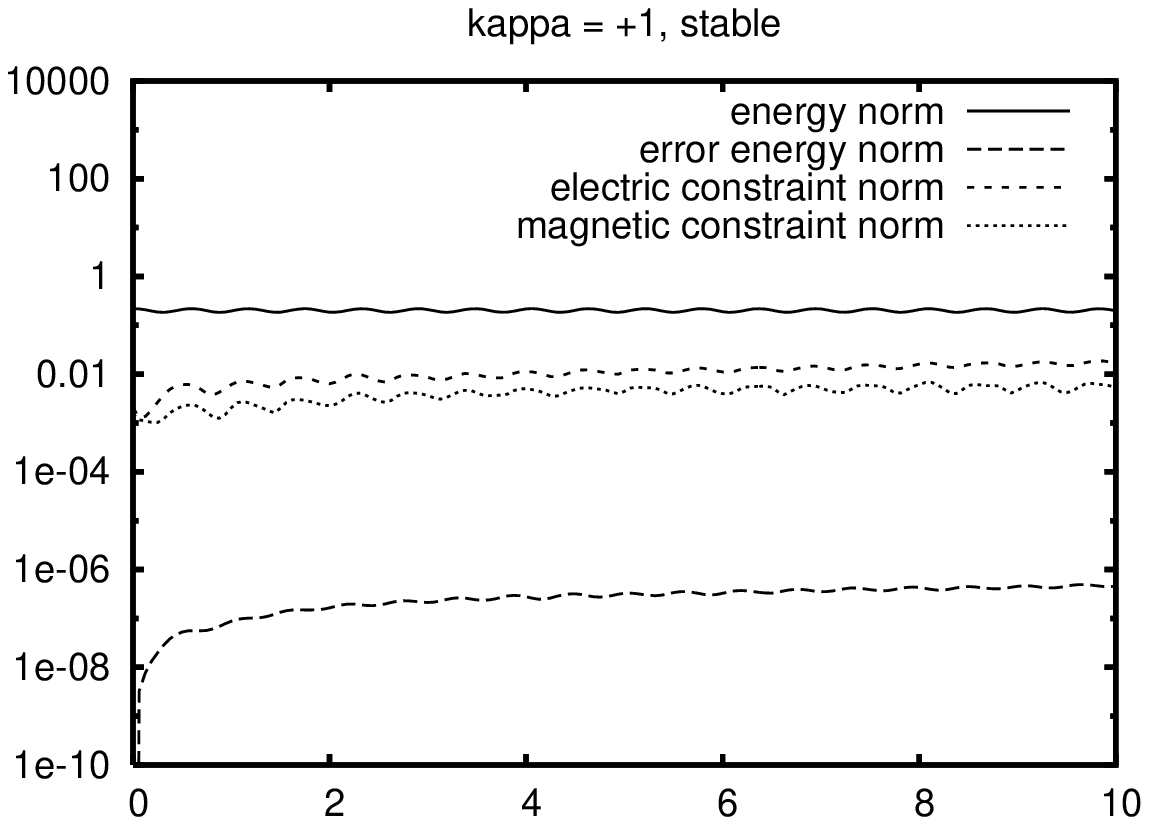}
   \hfill
   \includegraphics[width=0.49\textwidth]{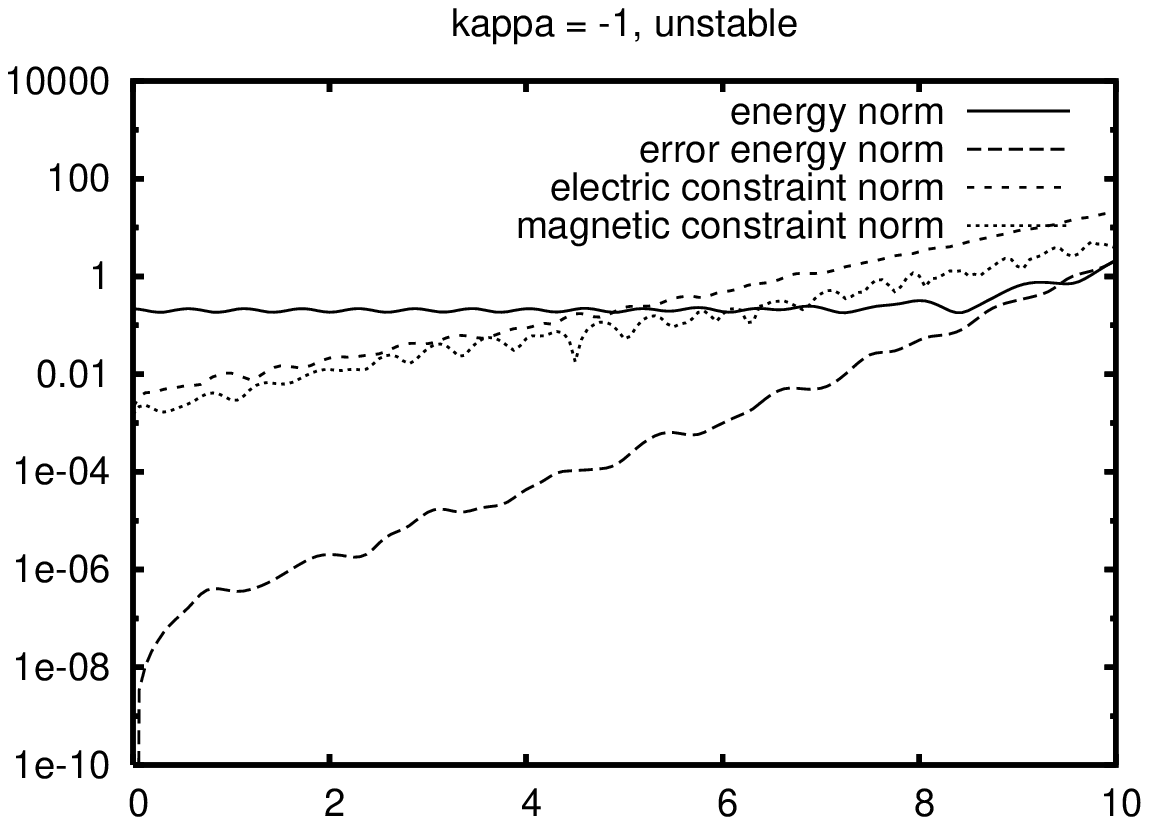}
  \caption{%
    (a) Evolution in stable foliation ($\kappa = +1$); behaves nicely,
    energy conserved, drift away from exact solution \emph{better}
    than linear. %
    (b) Evolution in unstable foliation ($\kappa = -1$), behaves very
    badly, exponential growth of both constraints, exponential
    deviation from exact solution, finally triggers exponential growth of the
    energy.%
    \label{fig:minkcomp2}
  }
\end{figure}

\pagebreak
In nonlinear situations, the decay of the fields is delayed by
nonlinear interactions, and the ODE effects have an even stronger
influence than in linear situations. In order to monitor these effects
over a considerable amount of time, we consider a finite box with
ideally conducting walls, i.e. a cavity in which the field excitation
is reflected back and forth. We foliate Minkowski space with simple
hyperboloids that are bent only in x-direction and are flat in
yz-directions: $\{t = \const\}$-surfaces with $t(T,X,Y,Z) = T -
\kappa (\sqrt{1+X^2} - 1)$.  For initial data we use analytically
known eigenmodes of the cavity, transformed appropriately from
standard Minkowski to the curved coordinates. The results of our
experiments are presented in figs. \ref{fig:minkcomp1} and \ref{fig:minkcomp2}.

\section{Solution of the constraints}
Solving the constraints is interesting from two perspectives:
first it is a necessary prerequisite for evolutions, and second, since
a general procedure for solving the regular conformal constraints is not known,
it provides an interesting example of a more ad-hoc regularization procedure
for the Einstein equations.
We consider an isotropic initial hypersurface, i.e. 
$\tilde{\chi}_{ab} = \tilde{\chi}\tilde{h}_{ab}/3$ with 
$\tilde{\chi}=\const$. This ansatz solves the momentum constraint
and is in some sense analogous to time symmetry for asymptotically euclidean slices.
By applying the Lichnerowicz-York procedure to the \textit{rescaled} metric 
$\Omega^2\tilde{h}_{\mu\nu} = \phi^4h_{\mu\nu}$, the Hamiltonian constraint is converted 
into the Yamabe equation
$$
4\Omega^2D^\mu D_\mu\phi - 4\Omega\:D^\mu\Omega\:D_\mu\phi
- \left(\frac{R}{2}\Omega^2 + 2\Omega\:D^\mu D_\mu\Omega
        - 3\:D^\mu\Omega\:D_\mu\Omega\right)\phi
\;=\; \frac{1}{3}\tilde{\chi}^2\phi^5\;,
$$
where $D_\mu$ denotes the spatial covariant derivative operator
and $R$ its Ricci scalar. For $\Omega\neq 0$ this is a semilinear elliptic equation,
but its principal part vanishes on the conformal boundary 
and standard elliptic theory cannot be applied.
The existence of smooth solutions $\phi$ has been proven in \cite{ACF} 
under the condition that the extrinsic 2-curvature induced on the initial cut of 
$\scri$ by the \emph{free metric} is pure trace.
The Yamabe equation then also determines the boundary values to be
$\phi^2 = 3|\tilde{\chi}|^{-1}\sqrt{D^\mu\Omega\:D_\mu\Omega}\;$ on $\scri$.

As an example, we consider the simple \emph{axisymmetric} Brill ansatz
$$
d\sigma^2 = e^{aq(\rho,z)}\left(d\rho^2 + dz^2\right) + \rho^2 d\varphi^2,\quad
q(\rho,z) = \rho^2e^{-(\rho^2+z^2)}.
$$
Such data are well studied in the asymptotically euclidean regime where it is known
that for small amplitudes $a$ the waves eventually disperse, leaving flat space
behind, whereas for large values of $a$ the waves collapse and form trapped surfaces
(in particular we have used such data to test our code
against known results \cite{Alcubierre:HorizonFinders}).
In the hyperboloidal case, the problem becomes nonlinear due to the non-vanishing of
$\tilde{\chi}$, which we set to unity without restricting generality.
Choosing the conformal gauge as $\Omega = 1-r^2$ puts $\Scri$ to $r=1$ and makes the
regularity condition on the extrinsic 2-curvature of $\Scri$ be identically satisfied.
The resulting nonlinear boundary value problem can be simply discretized with 
$2^{nd}$ order finite differences and solved through a preconditioned GMRES method 
\cite{Saad}.

\begin{figure}[htbp]
  \centering
  \PSFig{0.49\textwidth}{!}{270}{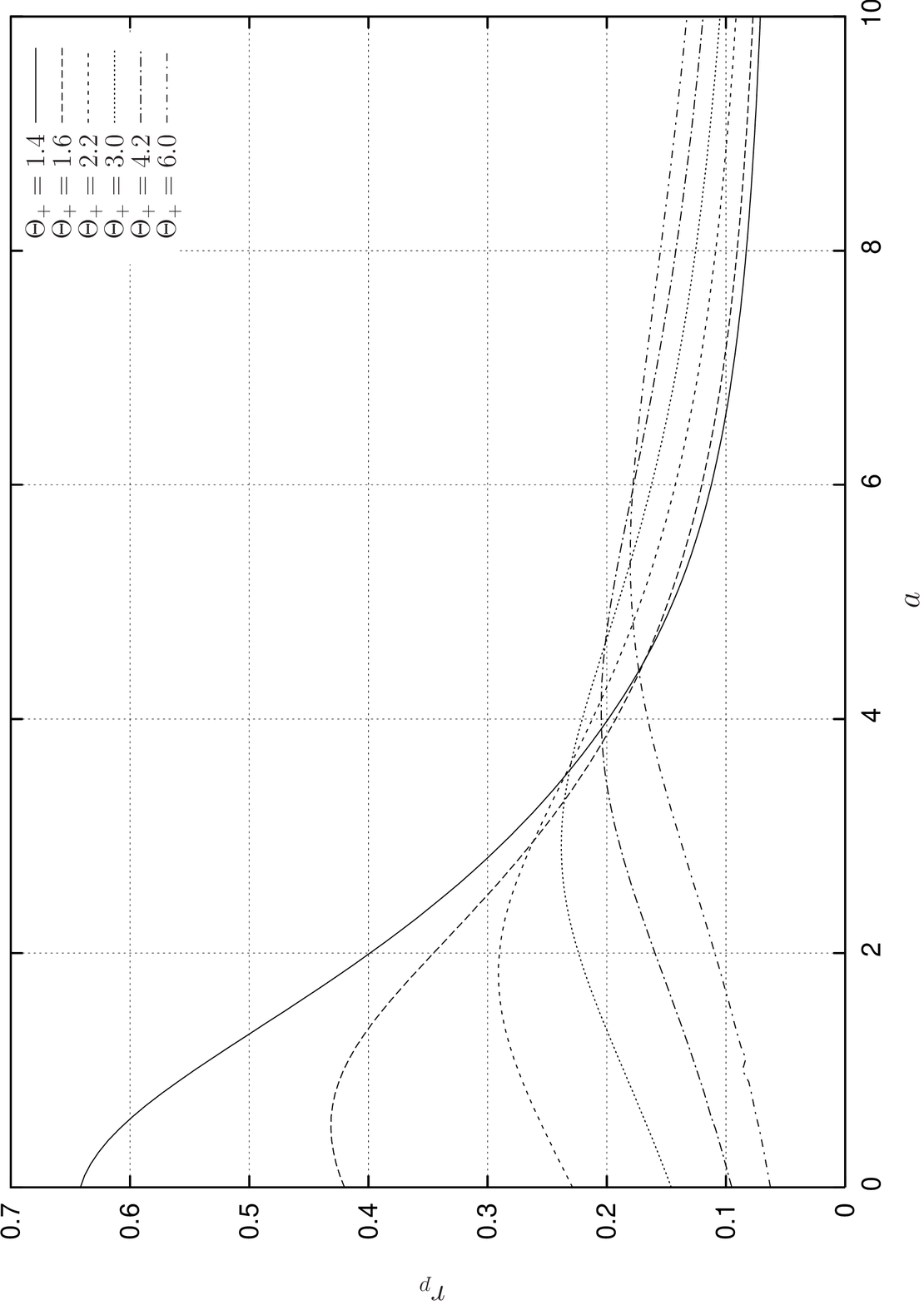}
  \hfill
  \PSFig{0.49\textwidth}{!}{270}{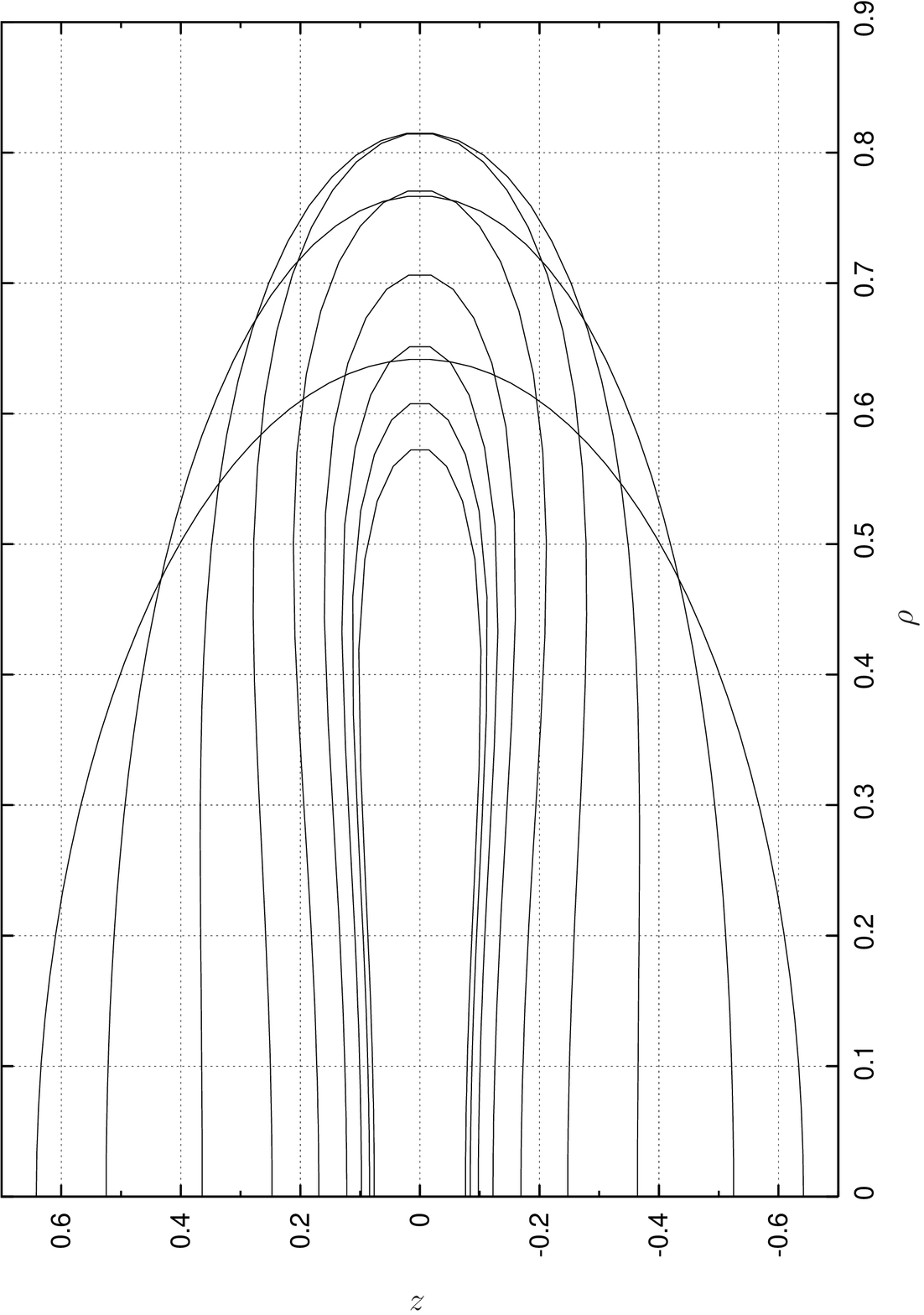}
  \caption{Left: Polar radii of equiexpansion surfaces for different amplitudes. Right: Shape of the surface $\Theta_+ = 1.4$ for $a=0$ (innermost), $3, 6, \ldots, 24$ (outermost). \label{fig:id}}
\end{figure}
For the physical interpretation of the data it is interesting to search for marginally 
trapped surfaces, i.e. surfaces on which the null expansion $\Theta_+$ vanishes.
Note that since $\Scri^+$ is a ``surface at infinity'', the expansions take the unique
values $\Theta_+ = \frac{4}{3}\tilde{\chi}$, $\Theta_- = 0$ there.
Since the geometry is euclidean in the vicinity of the axis,
the expansions have their flat space behavior 
$\Theta_\pm\to\pm\infty$ for $r\to 0$.
Marginal surfaces can now develop if there exist values 
of the amplitude $a$ for which $\Theta_+$ becomes non-positive in between. 
Surprisingly, while in the asymptotically euclidean case this happens generically,
for the classes of data we have studied, $\Theta_+$ remains strictly positive and no 
trapped surfaces exist even for extremely high amplitudes.

\section{Conclusions}

The prime motivation to study evolutions based on hyperboloidal slicings 
is that they enable us to reach null infinity with the flexibility of Cauchy codes. Using 
the example of the Maxwell equations we have discussed that hyperboloidal slices 
may tend to create either strong constraint damping or growth, which makes them interesting both 
as a model for what can go wrong and as a potential remedy. 
The general conformal field equations allow us to treat null and spacelike infinity in 
a unified picture, which we hope to help understand the physical significance of 
the idealizations one makes when using the compactified picture. 
In order to develop hyperboloidal codes that can handle physically interesting 
situations involving dynamical black holes and gravitational radiation, 
we believe it will be fruitful to obtain a fresh perspective on the compactification 
problem and consider adapted gauges as a starting point for regularizing
equations rather than proceeding in the opposite direction. 

%%%%%%%%%%%%%%%%%%%%%%%%%%%%%%%%%%%%%%%%%%%%%%%%
%% BACKMATTER
%%%%%%%%%%%%%%%%%%%%%%%%%%%%%%%%%%%%%%%%%%%%%%%%

\begin{acknowledgments}
  The authors would like to thank H. Friedrich for many useful 
  discussions, C. Lechner for her work on computer algebra tools,
  especially those for the Maxwell equations, and I. Hinder for his
  continuing development work on {\tt Kranc}. This work was supported in
  part by the SFB/Transregio 7 ``Gravitational Wave Astronomy'' of the
  German Science Foundation.
\end{acknowledgments}

\bibliographystyle{urlgerabbrv}   % if natbib is available
%\bibliography{ERE05}

\end{document}